\documentstyle[12pt,epsf]{article}
\newlength{\overeqskip}
\newlength{\undereqskip}
\newcommand{\nc}{\newcommand}

\nc{\be}[1]{\begin{equation} \mbox{$\label{#1}$}}
\nc{\bea}[1]{\begin{eqnarray} \mbox{$\label{#1}$}}
\nc{\Label}[1]{\label{#1}}
\nc{\bi}[1]{\bibitem{#1}}

\nc{\ee}{\end{equation}}
\nc{\eea}{\end{eqnarray}}
\nc{\sss}{\scriptscriptstyle}
\nc{\lsim}{\mbox{\raisebox{-.6ex}{~$\stackrel{<}{\sim}$~}}}
\nc{\gsim}{\mbox{\raisebox{-.6ex}{~$\stackrel{>}{\sim}$~}}}
\nc{\nn}{\nonumber}
\begin{document}
%
%
\begin{titlepage}
\pagestyle{empty}
\baselineskip=21pt
\rightline{Report No. UNIL-IPT/00-04}
%
\vskip 0.8in

\begin{center} 
  {\Large\bf A comment on baryogenesis at the electroweak scale\\
                    in alternative cosmologies}
\end{center}
\vskip .3in

\begin{center}
     {\Large Tomislav Prokopec}\\
\vskip 0.1 in
      {\it Universit\'e de Lausanne, Institut de Physique Th\'eorique,
BSP, CH-1015 Lausanne, Suisse}\\
{\it E-mail: Tomislav.Prokopec@ipt.unil.ch}
\end{center}
\vskip 0.3 in

\centerline{ {\bf Abstract} }
\baselineskip=18pt
\vskip 0.5truecm
\noindent

In this study we examine post-inflationary cosmologies dominated by a scalar
field with the equation of state 
$p_\phi=w_\phi \rho_\phi$ $\,(0\leq w_\phi\leq 1)$
in order to facilitate baryogenesis at the electroweak scale. We take a more
conventional approach from one in Ref.~\cite{jp} and assume that 
the Universe reheats by the scalar field decay before the nucleosynthesis 
epoch, and find a larger expansion rate at the electroweak scale than the one
obtained in Ref.~\cite{jp}. 
The decaying field models suffer however from an entropy release that dilutes
the baryon number produced at the electroweak scale. This dilution is
minimized when the kinetic scalar field mode dominates ($w_\phi = 1$),
singling it out as the preferred cosmology with regard to baryogenesis.
We study both cases, the electroweak transition with an expansion driven
departure from equilibrium, and a first order phase transition. We show that
in the former case with some tuning one can produce the amount of matter
consistent with observation. In the latter case the expansion rate at the
electroweak scale may be almost as large as the symmetric phase sphaleron 
rate, so that even the electroweak models with a relatively weak first order 
phase transition can be viable for baryogenesis. 

\end{titlepage}

\baselineskip=20pt

%
%
\section{Introduction}

The main difficulty facing almost any electroweak baryogenesis scenario is 
the requirement for a strong electroweak phase transition. This is required 
in order to prevent wash out of the produced baryons by
the subsequent sphaleron transitions, and it is expressed in the (in)famous 
sphaleron bound. The bound states roughly that the jump in the 
order parameter (the Higgs expectation value) must be
greater then the temperature of the phase transition \cite{revw}. 
This requirement has already killed the Minimal Standard Model (MSM) as 
a candidate for electroweak baryogenesis. Consequently baryogenesis efforts
have been redirected toward the extensions of the Standard
Model which provide sufficiently strong first order phase transition,
most popular one being the Minimal Supersymmetric Standard Model. 
Such efforts are very natural, and yet they do not take account of 
the very basic fact: we do not have any direct observational constraints on
the Universe before the nucleosynthesis epoch.
In this letter following Ref.~\cite{jp} we explore this simple fact 
and discuss alternative cosmologies that predate the nucleosynthesis epoch.

 Let us first however overview some of the basic facts of the electroweak
scale baryogenesis. At the electroweak transition in the presence of a 
chemical potential for baryon number that is biased by some CP-violating 
source, quite generically one gets for the baryon-to-entropy ratio 
\begin{equation}
\frac{n_B}{s}\sim \frac{\delta_{\rm CP}}{g_*}
\left(\frac{H}{T}\right)_{\rm freeze}\,,
\label{kin.2pt.1}
\end{equation}
where $\delta_{\rm CP}$ is the effective CP-violating parameter, and
$g_*$ is the number of relativistic degrees od freedom, which in the MSM
reads $g_*\simeq 106.75$.  The expansion rate $H_{\rm freeze}$ is defined at
the temperature $T_{\rm freeze}$ at which the sphaleron processes
go out of equilibrium (freeze out). For our purposes they can be 
identified with their values at the electroweak scale, since
$T_{\rm freeze}\approx T_{\rm ew}$ and $H_{\rm freeze}\approx H_{\rm ew}$. 
A derivation of equation~(\ref{kin.2pt.1}) in the two Higgs doublet model can 
be found in the first reference of~\cite{jp}, Eq.~(44). In this case the
effective CP-violation parameter
$\delta_{\rm CP}\simeq - (Td\chi/dT)_{\rm freeze}$,
where $d\chi/dt$ is the chemical potential for the baryon number in the two 
Higgs doublet model. We have shown that quite naturally, 
$(Td\chi/dT)_{\rm freeze}$ may be of the order unity, and with some amount of 
tuning it can be as large as $10^2$. Recall that in the standard radiation
dominated cosmology the Hubble parameter at the electroweak scale equals
\begin{equation}
\frac{H_{\rm ew}}{T}\simeq 1.4 \times 10^{-16}
  \left(\frac{g_*}{107}\right)^\frac{1}{2}
  \left(\frac{T}{100{\rm GeV}}\right) .
\label{kin.2pt.2}
\end{equation}
This then implies that the observed amount of matter in the Universe,
as given by the nucleosynthesis constraint~\cite{bt}
\begin{equation}
\left(\frac{n_B}{s}\right)_{\rm obs}\sim 4-10\times 10^{-11}
\label{kin.2pt.3}
\end{equation}
is inconsistent with Eqs.~(\ref{kin.2pt.1}) and~(\ref{kin.2pt.2}).
Because of this simple result attempts to produce baryons at the electroweak
transition, when the main source for departure from 
equilibrium is expansion driven, have faded.

\section{The model ..}

We will now show how this inconsistency can be avoided by 
considering the following simple model. Let us assume that the  
post-inflationary Universe is dominated by a scalar field $\phi$.
(This field may or may not be the inflaton.) We assume that $\phi$ decays
into particles, which then instantly  thermalize. 
This situation is realized for the following hierarchy of couplings 
\begin{equation}
\Gamma_\phi\ll H\ll \Gamma_{\rm th} ,
\label{kin.2pt.4}
\end{equation}
where $\Gamma_\phi$ is the decay rate of $\phi$, and $\Gamma_{\rm th}$ the
thermalization rate of the decaying products. 
We shall assume that $\phi$ obeys the following equation of state
\begin{equation}
p_\phi=w_\phi\rho_\phi, \qquad 0\leq w_\phi\leq 1 . 
\label{kin.2pt.5}
\end{equation}
With this, the relevant equations of motion are simply
\begin{eqnarray}
\frac{d\rho_\phi}{dt}+nH\rho_\phi+\Gamma_\phi(\rho_\phi-\rho_\phi^{\rm eq}) 
  &=& 0 , \qquad n=3(w_\phi+1)
\label{kin.2pt.6}
\\
\frac{d\rho_r}{dt}+4H\rho_r-\Gamma_\phi(\rho_\phi-\rho_\phi^{\rm eq}) 
  &=& 0 ,
\label{kin.2pt.7}
\end{eqnarray}
where $H$ is the Hubble parameter given by 
\begin{equation}
H^2 = \frac{\rho_\phi+\rho_r}{3M_P^3} ,
\label{kin.2pt.8}
\end{equation}
$M_P=(8\pi G)^{-1/2}\simeq 2.4\times 10^{18}$GeV is the reduced Planck mass,
and $\rho_\phi$ and $\rho_r$ are the energy densities of the scalar field and 
radiation fluid, respectively. In writing Eqs.~(\ref{kin.2pt.6})
and~(\ref{kin.2pt.7}) we used 
$\rho_\phi-\rho_\phi^{\rm eq}=\rho_r-\rho_r^{\rm eq}$, and 
$\Gamma_\phi=\Gamma_{\phi\rightarrow\rm rad}+
\Gamma_{\rm rad\rightarrow \phi}$ is the sum of the scalar field decay rate
and the (inverse) re-population rate. Provided the hierarchy~(\ref{kin.2pt.4}) holds, this simple model of inflaton 
decay is a good description not only when perturbative decays dominate (old 
reheating theory), but also when the field $\phi$ decays non-perturbatively
{\it via} parametric resonance (modern reheating theory). This is so because
the exponential enhancement in the decay rate characterizing parametric 
resonance is not operative when thermalization rate of the decaying 
products is very large. The equilibrium densities are related as
$\rho_{\rm tot}\equiv\rho_\phi+\rho_r=\rho_\phi^{\rm eq}+\rho_r^{\rm eq}$, and 
$\rho_\phi^{\rm eq}/g_\phi=\rho_r^{\rm eq}/ g_*$, where 
$g_\phi$ is the number of degrees of freedom in $\phi$. In order to keep
Eq.~(\ref{kin.2pt.5}) more general, we have left the ``damping'' coefficient
$n=3(w_\phi+1)$ unspecified, so that for example $n=3$ $(n=4)$ corresponds to a
massive (massless) field, and $n=6$ corresponds to a scalar field dominated
by the kinetic energy (kination). As discussed in Ref.~\cite{jp} a simple
realization of kination is the following exponential potential
\begin{equation}
V(\phi)=V_0e^{ -\lambda\phi/M_P},\qquad 
  V_0=\frac{2}{\lambda^2}\left(\frac{6}{\lambda^2}-1\right)M_P^4 .
\label{kin.2pt.9}
\end{equation}
For $\lambda^2<6$ there is an attractor solution of the form
\begin{equation}
\phi(t)=\frac{2M_P}{\lambda}\ln M_Pt,\qquad
a\propto t^{2/\lambda^2} ,
\label{kin.2pt.10}
\end{equation}
so that in the limit when 
$ \lambda\rightarrow \sqrt {6}$ one obtains kination with 
\begin{equation}
  \rho_\phi = \rho_0\left(\frac{a_0}{a}\right)^6 ,\qquad 
a\propto t^{1/3} .
\label{kin.2pt.11}
\end{equation}
This case corresponds to $n=6$ ($w_\phi=1$) in Eq.~(\ref{kin.2pt.6})
a behaviour identical to one with a completely flat potential.  
In the opposite limit, when $\lambda^2>6$, one gets inflation.

\section{.. its solution ..}

Equations~(\ref{kin.2pt.6}) and~(\ref{kin.2pt.7}) can be easily 
solved in the limit when only a small fraction of $\phi$ has decayed into
radiation, $\rho_\phi\gg \rho_r,\rho_\phi^{\rm eq}$. The result is 
({\it cf.} Ref.~\cite{gpr} and~\cite{kt})
\begin{eqnarray}
\rho_\phi  &=& \rho_\phi^{\rm eq} + \rho_0\left(\frac{a_0}{a}\right)^n
    e^{-\Gamma_\phi t}
\label{kin.2pt.12}
\\
\rho_r  &=& \Gamma_\phi\rho_0\left(\frac{a_0}{a}\right)^4 \int_{t_0}^t dt
           \left(\frac{a_0}{a}\right)^{n-4} e^{-\Gamma_\phi t} .
\label{kin.2pt.13}
\end{eqnarray}
 Since we are primarily interested in integrating this equation during 
the preheating epoch when $\rho_\phi\gg \rho_r$, $\Gamma_\phi t\ll 1$,
and $a\propto t^{2/n}$, we obtain
\begin{eqnarray}
\rho_r  &=& \frac{2}{8-n}\frac{\Gamma_\phi}{H_0}\rho_0
  \left(\frac{t_0}{t} - \left(\frac{t_0}{t} \right)^\frac{8}{n} \right)
 \left(1+o(\Gamma_\phi t)\right) 
\nonumber\\
&=& \frac{\pi^2}{30}\,g_* T^4 ,
\label{kin.2pt.14}
\end{eqnarray}
where $t_0=2/nH_0$, $H_0^2=\rho_0/3M_P^2$. Note that $\rho_r$ (and $T$) grows
rapidly from zero, reaching quickly 
(at $t_{\rm max}\simeq t_0 (8/n)^{n/(8-n)}$) a maximum value 
$\rho_{r\,\rm max}\simeq \rho_0(\Gamma_\phi/4H_0)(n/8)^{n/(8-n)}$,
after which $\rho_r\propto t^{-1}$ ($T\propto t^{-1/4}\propto a^{-n/8}$).
This dependence continues until the reheating temperature $T_{\rm reh}$
is reached, when $\phi$ and radiation begin to equilibrate:
$\rho_\phi/g_\phi \sim \rho_r/g_*$.
Quite generically this occurs when $\Gamma_\phi t\sim 1$. At that moment 
the solutions~(\ref{kin.2pt.12}) -- (\ref{kin.2pt.14}) break down, and 
the Universe enters radiation era. 

\section{.. and the consequences} 

\subsection{On the expansion rate}

From Eq.~(\ref{kin.2pt.14}) we infer the following expression for
the expansion rate 
\begin{eqnarray}
H \simeq \frac{2}{nt} 
= \frac{8-n}{6}\,\frac{\rho_r}{\Gamma_\phi M_P^2}
\propto T^4 ,\qquad 
(T_{\rm max}> T > T_{\rm reh}) .
\label{kin.2pt.15}
\end{eqnarray}
This dependence of the expansion rate on the temperature is generic in that
it does not depend on $w_\phi=(n-3)/3$ in the equation of state for
$\phi$~(\ref{kin.2pt.5}). This type of behaviour is precisely the desired one
since it may result in a large expansion rate at the electroweak scale. 
Since at the nucleosynthesis epoch the Universe is constrained to be
radiation dominated, we must have $T_{\rm reh} > T_{\rm ns}\sim 2$MeV.
This means that, in comparison to the 
standard Hubble parameter at the electroweak scale~(\ref{kin.2pt.2}),
the Hubble parameter in Eq.~(\ref{kin.2pt.15}) is enhanced as 
\begin{eqnarray}
\frac{H}{T_{\rm ew}} \simeq \frac{H_{\rm ew}}{T_{\rm ew}}
 \left(\frac{T_{\rm ew}}{T_{\rm reh}}\right)^2
,
\label{kin.2pt.16}
\end{eqnarray}
so that, when the reheat temperature is tuned to be 
$T_{\rm reh}\sim T_{\rm ns}$, the expansion rate at the electroweak scale 
may be as much as $(T_{\rm ew}/T_{\rm ns})^2\sim 10^{10}$ times larger from 
the standard one~(\ref{kin.2pt.2}). Note that with this much enhancement
in the expansion rate the computed baryon number production at the electroweak 
scale~(\ref{kin.2pt.1}) and the observed value~(\ref{kin.2pt.3}) are 
consistent. 

In order to check consistency of our model, we need to make sure 
that the sphaleron rate in the symmetric phase (above the electroweak scale)
is still in equilibrium with the enhanced expansion rate~(\ref{kin.2pt.16}).
The maximum  attainable rate at the electroweak scale is about 
$H_{\rm ew\,max}/T \sim 10^{-6}$. On the other hand, the rate equation
for the baryon number density $n_B$ is given by \cite{am}
\begin{equation}
\frac{dn_B}{dt}+3Hn_B +\frac{13N_F}{4} \frac{\bar\Gamma_s}{T^3} n_B =0
\label{kin.2pt.17}
\end{equation}
where $N_F= 3$ is the number of fermion families.
In the symmetric phase the sphaleron rate reads
$\bar\Gamma_{\rm s}\simeq 25\pm 2\; \alpha_w^5 T^4$ \cite{gm}, and
$\alpha_w=g^2/4\pi\simeq 1/29$ is the strength of the weak coupling at the
electroweak scale. This then implies that baryons are destroyed at the rate
$\Gamma_{\rm sph}\sim 10^{-5} T\gg H_{\rm ew\, max}\sim 10^{-6}T$,
which was an underlying assumption \cite{jp} made in estimating the
baryon-to-entropy ratio~(\ref{kin.2pt.1}). In this case the sphalerons are out
of equilibrium above about $T\sim 10^3$GeV, while other species in the plasma
typically fall out of equilibrium when $T> 10^4 - 10^5$GeV. An exception
is the right handed electron, whose equilibration rate is of the order 
$\Gamma_{e_R}\sim  10^{-12}T$, and hence may be out of
equilibrium at the electroweak scale. This may have interesting consequences 
worth further investigation.

\subsection{On the baryon number dilution}
\label{sec: Baryon number dilution}

 We have so far shown that one can produce enough baryons at the electroweak
scale, but we have not taken account of the dilution in the baryon-to-entropy
rate caused by the entropy released by the decay of $\phi$.
This can be estimated as follows. After the sphalerons freeze out, 
the number of baryons per comoving volume, $a^3 n_B$, remains constant. 
On the other hand the entropy per comoving volume scales as 
$S_{\rm com}\equiv a^3 s\propto a^3 T^3\propto T^{-3(8-n)/n}$,
where we made use of $a\propto t^{2/n}\propto T^{-8/n}$,
so that, when the entropy dilution is included, we get the following  
estimate for the baryon to entropy ratio that survives
\begin{eqnarray}
\frac{n_B}{s} &\sim&
 \left(\frac{n_B}{s}\right)_{\rm produced}
   \left(\frac{T_{\rm reh}}{T_{\rm ew}}\right)^\frac{3(8-n)}{n} 
\nonumber\\
 &\simeq& \frac{\delta_{\rm CP}}{g_*}
  \frac{H_{\rm ew}}{T_{\rm ew}}
   \left(\frac{T_{\rm ew}}{T_{\rm reh}}\right)^\frac{5n-24}{n} .
\label{kin.2pt.18}
\end{eqnarray}
Note that for $n> 24/5$ ($w_\phi>3/5$) one obtains a net enhancement in 
the baryon production in comparison to the estimate~(\ref{kin.2pt.1}).
In particular, when $n=6$ (kination) the enhancement is
$T_{\rm ew}/T_{\rm reh}$, which can be as large as $10^5$,
so that, in order to get a baryon production consistent with 
the observation~(\ref{kin.2pt.3}), it is required that the 
effective CP-violating parameter $\delta_{\rm CP}$ be at least of the order 
$10^2$. This is possible to achieve with a certain amount of tuning,
as explained on the example of the two Higgs doublet model in Ref.~\cite{jp}.
The main difference is that in Ref.~\cite{jp} we assumed 
an unconventional reheating as first discussed by Spokoiny in Ref.~\cite{s},
while in this letter we have assumed a more standard type of reheating 
where the inflaton is weakly coupled to matter, so that it decays very
slowly either perturbatively or nonperturbatively {\it via}
narrow parametric resonance. To sum up, we have found that baryon production
is quite generically enhanced in models in which inflation is followed
by a kinetic mode domination, irrespectively on whether the Universe reheats by
the inflaton decay or by a nonstandard Spokoiny reheating mechanism, 
provided reheating ends around (but not later than) the nucleosynthesis epoch.

 The question is of course whether our cosmological model can be made 
consistent with all observational constraints. A first guess would be
{\it yes,} simply because we have explicitly constructed such a model 
in Ref.~\cite{jp} with one less free parameter ($\Gamma_\phi$).
Let us nevertheless show that this is indeed the case. 
First we have  $\rho_{\rm reh} \sim (\Gamma_\phi M_P)^2 \sim T_{\rm reh}^4 
>  T_{\rm ns}^4 \sim ({\rm MeV})^4$,
from which we conclude that decay rate is tiny
\begin{equation}
\Gamma_\phi \sim \frac{T_{\rm reh}^2}{M_P} > \frac{T_{\rm ns}^2}{M_P}
\sim 10^{-12}\,{\rm eV} .
\label{kin.2pt.19}
\end{equation}
Further, from
\begin{equation}
 \frac{H_0}{\Gamma_\phi}
 \sim \left(\frac{T_{\rm max}}{T_{\rm reh}}\right)^4 \gg
 \left(\frac{T_{\rm ew}}{T_{\rm ns}}\right)^4 \sim 10^{20} ,
\label{kin.2pt.20}
\end{equation}
we obtain the following constraint on the expansion rate of the 
Universe at the end of inflation 
\begin{equation}
H_0\sim \frac{T_{\rm max}^4}{M_PT_{\rm reh}^2}
   > \frac{T_{\rm ew}^4}{M_PT_{\rm ns}^2} \sim 10^{-5}\,{\rm GeV} .
\label{kin.2pt.21}
\end{equation}
These two constraints are very mild in regard to constraining inflationary
models. It is now easy to check that, even for the minimum
decay rate $\Gamma_\phi\sim 10^{-12}$eV consistent with Eq.~(\ref{kin.2pt.19}),
the Universe reheats by the inflaton decay, and not by the Spokoiny mechanism.
Indeed from
\begin{equation}
 T_{\rm max}^4\sim H_0\Gamma_\phi M_P^2
 \; \gg \;T_{\rm Hawking}^4 \sim \left(\frac{H_0}{2\pi}\right)^4 
\label{kin.2pt.22}
\end{equation}
we infer 
\begin{equation}
\Gamma_\phi\gg \frac{H_0^3}{M_P^2}\, >\, 10^{-43}{\rm eV} ,
\label{kin.2pt.23}
\end{equation}
so that the Spokoiny reheating mechanism becomes operational only for 
the miniscule decay rates for which this bound is violated.

One would think that it is necessary to impose the COBE constraints on the 
amplitude of cosmological perturbations and their spectral index as well.
This is however not the case, since they actually constrain inflationary
potentials, and hence should be imposed once a particular realization of
inflation is given. Since the considerations in this letter are to a large
extent generic, we need not consider the COBE constraints here. 
Finally we point out that, for most inflationary models considered in
literature \cite{lr}, the COBE constraints are in concordance with
Eqs.~(\ref{kin.2pt.19}) and~(\ref{kin.2pt.21}).

\subsection{On the sphaleron bound}

 We now briefly discuss the relevance of our alternative cosmologies 
for baryogenesis at a first order electroweak phase transition. 
In order to avoid the sphaleron erasure the following bound has
to be satisfied \cite{revw}
\begin{equation}
 \frac{v(T_{\rm ew})}{T_{\rm ew}} > -b\ln\frac{H}{T_{\rm ew}}\, ,
\label{kin.2pt.24}
\end{equation}
where $v$ denotes the jump in the order parameter
(the Higgs expectation value) at the phase transition, and $b$ is a weak 
function of $H$ (or equivalently $v$) and the Higgs mass. For a rough 
estimate we may set it to a constant ($b\sim 0.03$), such that for the
standard expansion rate~(\ref{kin.2pt.2}) one obtains $v\gsim 1.1 T_{\rm ew}$. 
With the expansion rate~(\ref{kin.2pt.15}), the bound gets modified as 
\begin{equation}
 \frac{v (T_{\rm ew})}{T_{\rm ew}} > -b\ln\frac{H_{\rm ew}}{T_{\rm ew}} 
 -2b\ln \frac{T_{\rm ew}}{T_{\rm reh}}\, .
\label{kin.2pt.25}
\end{equation}
Now when $T_{\rm ew}/T_{\rm reh}\sim 10^5$, for which in the symmetric 
phase $\Gamma_{\rm sph}\sim 10 H_{\rm max}$, the sphaleron bound becomes 
largely relaxed, yielding $v\gsim 0.4T_{\rm ew}$. Following 
Refs.~\cite{am,jp,m2} we have performed a more careful computation and 
obtained an almost identical estimate. The reader should be 
alert to that these estimates assume validity of the perturbative expression
for the sphaleron rate in the region where it is barely trustable
($\alpha_3\sim 0.15$ in figure 3 of Ref.~\cite{am}). To get a more precise 
estimate of the sphaleron bound in the weak transition regime would require
numerical estimation. 

To conclude, we have shown that, when $T_{\rm reh}\gsim T_{\rm ns}$
the sphalerons freeze out at the transition even when the transition is
quite weak ($v\sim 0.5T$). (They remain frozen after the
transition simply because the sphaleron rate drops faster then the expansion
rate~(\ref{kin.2pt.15})). This of course implies that, in cosmologies
with a kinetic scalar field mode domination, in many cases the sphaleron bound
does not affect baryon production at a first order
electroweak transition. Needless to say this opens a new window for 
baryogenesis scenarios operative in models 
that result in a weakly first order electroweak phase transition.
We must not however forget the entropy release from the decay of $\phi$,
which we consider next. 

Now following the reasoning at the beginning of
section~\ref{sec: Baryon number dilution}, we infer that the
baryon-to-entropy ratio that survives today can be written as
({\it cf.} Eq.~(\ref{kin.2pt.18}))
\begin{equation}
\left(\frac{n_B}{s}\right)_{\rm today}
 \sim \left(\frac{n_B}{s}\right)_{\rm created}
   \left(\frac{T_{\rm ew}}{T_{\rm reh}}\right)^\frac{3(8-n)}{n} .
\label{kin.2pt.26}
\end{equation}
Note that the dilution of the produced baryon-to-entropy ratio is large in
models with conventional reheating ($n=3,4$), while it is relatively small
in the case of kination ($n = 6$). More explicitly, for a massive field 
$\phi$ ($n=3$) the entropy dilution factor is $(T_{\rm ew}/T_{\rm reh})^5$,
which for $T_{\rm ew}/T_{\rm reh}\sim 10^5$ can be as large as $10^{25}$. 
This case has been discussed in a recent preprint~\cite{dlr}. For 
a massless field $\phi$ ($n=4$) the situation is a little better: the dilution
factor is $(T_{\rm ew}/T_{\rm reh})^3<10^{15}$, which may still be very large.
For kination however ($n=6$), the dilution factor reads
$T_{\rm ew}/T_{\rm reh}< 10^5$, so that the required baryon-to-entropy ratio
produced at the electroweak scale 
is about $3-10\times 10^{-6}(T_{\rm ns}/T_{\rm reh})$, which may be 
produced by many electroweak scale baryogenesis mechanisms.  

\section{Conclusions}

 In this letter we have considered alternative cosmologies
in which, after an inflationary epoch, the Universe enters
a scalar field $\phi$ domination epoch with an atypical equation of state,
$p_\phi=w_\phi\rho_\phi$, $0\leq w_\phi\leq 1 $.
(This field may or may not be the inflaton.) We have then assumed that 
$\phi$ decays before the nucleosynthesis epoch, so that the standard 
nucleosynthesis is unaffected. When the decay rate $\Gamma_\phi$ is chosen
such that $\phi$ decays in between the electroweak and 
nucleosynthesis scales, beneficial consequences for baryogenesis incur,
analogous to ones discussed in Ref.~\cite{jp}. In this letter we have 
examined baryogenesis at the electroweak scale when the transition proceeds
({\bf a}) without a phase transition, or ({\bf b}) {\it via} a first order
phase transition. We have then shown that the most beneficial results in 
regard to electroweak baryogenesis are attained when the Universe is dominated
by the kinetic mode of the scalar field (kination), for which
$p_\phi=\rho_\phi$.

In Case ({\bf a}) we have found that the resulting enhancement in the 
baryon-to-entropy ratio is at most $T_{\rm ew}/T_{\rm reh}< 10^{5}$ ({\it cf.} 
Eq.~(\ref{kin.2pt.18})).  With some tuning in the CP-violating sector, 
this then may lead to a baryon number production consistent with
the observed value~(\ref{kin.2pt.3}).

In Case ({\bf b}) we have shown that quite generically (independent on 
the equation of state for $\phi$) the sphaleron bound can be 
relaxed as indicated in Eq.~(\ref{kin.2pt.25}).
The price to pay is a suppression of the original baryon-to-entropy ratio
produced at the electroweak transition due to the subsequent entropy release.
The model leading to a minimum entropy release is again kination 
($p_\phi=\rho_\phi$), when the produced baryon-to-entropy ratio
is diluted by $T_{\rm ew}/T_{\rm reh}< 10^5$. With this amount of dilution
many conventional baryogenesis mechanisms at the first order electroweak 
phase transition remain viable.

\section*{Acknowledgements}

I would like to thank Massimo Giovannini and Misha Shaposhnikov for useful
discussions. 

%
%
%
%
\nc{\AP}[3]    {{\it Ann.\ Phys.\ }{{\bf #1}, {#2} {(#3)}}}
\nc{\PL}[3]    {{\it Phys.\ Lett.\ } {{\bf #1}, {#2} {(#3)}}}
\nc{\PR}[3]    {{\it Phys.\ Rev.\ } {{\bf #1}, {#2} {(#3)}}}
\nc{\PRL}[3]   {{\it Phys.\ Rev.\ Lett.\ } {{\bf #1}, {#2} {(#3)}}}
\nc{\PREP}[3]  {{\it Phys.\ Rep.\ }{{\bf #1}, {#2} {(#3)}}}
\nc{\RMP}[3]   {{\it Rev.\ Mod.\ Phys.\ }{{\bf #1}, {#2} {(#3)}}}
\nc{\IBID}[3]  {{\it ibid.\ }{{\bf #1}, {#2} {(#3)}}}
%
%

\end{document}